\documentstyle[12pt,epsfig,here]{article}
\begin{document}
\newcommand{\beq}{\begin{equation}}
\newcommand{\eeq}{\end{equation}}
\def\beqn{\begin{eqnarray}}
\def\eeqn{\end{eqnarray}}

\newcommand{\Tr}{{\rm Tr}\,}
\newcommand{\E}{{\cal E}}

\newcommand{\ntwo}{${\cal N}=2\;$}
\newcommand{\none}{${\cal N}=1\;$}
\newcommand{\noneh}{${\cal N}=\,
^{\mbox{\small 1}}\!/\mbox{\small 2}\;$}
\newcommand{\vp}{\varphi}
\newcommand{\ve}{\varepsilon}
\newcommand{\pt}{\partial}


\begin{titlepage}
\renewcommand{\thefootnote}{\fnsymbol{footnote}}

\vfill

\begin{flushright}
FTPI-MINN-04/34, UMN-TH-2321/04\\
ITEP-TH-41/04
\end{flushright}


\begin{center}
\baselineskip20pt
{ \Large \bf   Spectral Degeneracy in Supersymmetric Gluodynamics and One-Flavor QCD related to \noneh SUSY}
\end{center}
\vfil
\begin{center}

{\large { \bf A.~Gorsky\,$^{a,b}$}} and 
{ \bf    M.~Shifman$^{b}$} 

\vspace{0.3cm}

$^a${\it Institute of Theoretical and Experimental Physics, Moscow
117259, Russia}\\
$^b${\it  William I. Fine Theoretical Physics Institute,
University of Minnesota,
Minneapolis, MN 55455, USA}

\vspace*{.45cm}

{\large\bf Abstract}
\end{center}
\vspace*{.05cm}

In supersymmetric gluodynamics (\none $\!\!\!$ super-Yang-Mills theory)
we show that the spectral functions induced by the {\em nonchiral}
operator $\Tr \left( G_{\alpha\beta}\, \bar\lambda^2\right)$
are fully degenerate in the $J^{PC}=1^{\pm -}$
channels. The above operator is related to
\noneh generalization of SUSY.
Using the planar equivalence, this translates into the statement
of    degeneracy between the mesons produced from the vacuum
 by the operators $
\left( \bar \Psi \vec E\Psi + i\bar \Psi \vec B \gamma^5\Psi
\right)
$
and
$
 \left( \bar \Psi  \vec B\Psi
- i
  \bar \Psi \vec E  \gamma^5\Psi\right)
$
in one-flavor QCD, up to $1/N$ corrections.
Here $\Psi$ is the quark field, and $\vec E ,\vec B$
are chromoelectric/chromomagnetic fields, respectively.

\end{titlepage}

Recently \cite{Oog,deBoer} a $C$-deformation of \none superalgebra
corresponding to non-anticommuting Grassmann coordinates $\theta$
which leads to the so-called \noneh superalgebra has been discussed. A $C$-deformed
supersymmetric (SUSY) Yang-Mills theory
(SUSY gluodynamics) was constructed and  discussed
by Seiberg \cite{Sei} (for  further developments
concerning some nonperturbative aspects of the theory see e.g. \cite{FD}).
The parameter of deformation $C$ in Ref.~\cite{Sei}
corresponds to the constant selfdual graviphoton background.

We will show that the very possibility of the above $C$-deformation
is related to a spectral degeneracy in {\em conventional} 
\none SUSY gluodynamics which, 
by virtue of the planar equivalence between SUSY and
orientifold theories \cite{ASV}, can be readily copied
in  one-flavor QCD, implying that the masses and coupling constants of hybrid $1^{\pm -}$ (color singlet) mesons are degenerate.
 That an infinite number of
spectral degeneracies must take place in the orientifold theory at
$N\to\infty$ was noted in~\cite{ASV}. The degeneracy we will discuss here is singled out because of its connection with a special operator
(of nonchiral type).

The above spectral degeneracy, in turn, suggests
that \noneh supersymmetry remains valid for coordinate-dependent
$C$ parameter.

The Lagrangian of  \none SUSY gluodynamics is
\beq
{\cal L}  =\frac{1}{2g^2}\int d^2\theta \, \Tr W^2 + \mbox{H.c.}\,,
\label{one}
\eeq
where 
\begin{equation}
{W}_{\alpha} = \frac{1}{8}\;\bar{D}^2\, \left( {e}^{-V}\! D_{\alpha } {e}^V\right) =
  i\left( \lambda_{\alpha} + i\theta_{\alpha}D - \theta^{\beta}\, 
G_{\alpha\beta} - 
i\theta^2{\cal D}_{\alpha\dot\alpha}\bar{\lambda}^{\dot\alpha} 
\right)\, . 
\label{sgfst}
\end{equation}
and $\int d^2\theta\,\theta^2 =1$.

The spectral degeneracy   in supersymmetric gluodynamics
follows from the fact that the $n$-point functions
\beqn
&&  \left\langle \Tr \left( G_{\alpha\beta} \bar\lambda^2 (x)\right) ,\,
 \Tr  \left( G_{\gamma\rho} \bar\lambda^2 (0)\right) \right\rangle_{\rm connected} = 0\,,
\nonumber\\[1mm]
&&...\nonumber\\[1mm]
&&\left\langle \Tr \left( G_{\alpha\beta} \bar\lambda^2 (x_n)\right) ,\, \Tr \left( G_{\alpha\beta} \bar\lambda^2 (x_{n-1})\right) ,\, ...\,,
 \Tr  \left( G_{\gamma\rho} \bar\lambda^2 (0)\right) \right\rangle_{\rm connected}
=0\,,
\label{two}
\eeqn
for any $n$ and $x_i$.
Here $\langle...\rangle$ stands for the vacuum expectation value 
($T$ product is implied on the left-hand side and elsewhere,
where necessary) and 
$$
\bar\lambda^2 \equiv \bar\lambda_{\dot\alpha}\bar\lambda^{\dot\alpha}
\,.
$$
Derivation of Eq.~(\ref{two}) is rather straightforward.
To begin, let us consider the two-point function in the first line.
We start from 
\beq
\{Q_\alpha ,\, \Tr\left( \lambda_\beta\, \bar\lambda^2\right)\} =
\Tr \left( G_{\alpha\beta}\, \bar\lambda^2\right),
\label{three}
\eeq
expressing the fact that the operator $\Tr \left( G_{\alpha\beta}\, \bar\lambda^2\right)$ is $Q$-exact.
Next, we replace one of the operators  $\Tr \left( G_{\alpha\beta}\, \bar\lambda^2\right)$ in the first line of
Eq.~(\ref{two}). Taking into account that $Q_\alpha |{\rm vac}\rangle =0$
we then transform it as follows:
\beq
 \left\langle\left\{
 \Tr \left( G_{\alpha\beta} \bar\lambda^2 (x)\right)  Q_\gamma\right\} ,\,
\Tr\left( \lambda_\rho\, \bar\lambda^2 (0)\right) \right\rangle\,.
\label{four}
\eeq
What remains to be done is to use the relation
\beq
\left\{ G_{\alpha\beta}Q_\gamma\right\}\sim {\cal D}_{\beta\dot\gamma}
\bar\lambda^{\dot\gamma}\,,
\label{five}
\eeq
which vanishes on mass shell. 
In other words, the operator $\Tr \left( G_{\alpha\beta}\, \bar\lambda^2\right)$ is $Q$-closed modulo the equation of motion.
In the correlation function
(\ref{four}) the equation of motion ${\cal D}_{\beta\dot\gamma}
\bar\lambda^{\dot\gamma}$ contracts
the propagator $\langle\bar\lambda (x)\, \lambda (0) \rangle$
resulting in the contact term
of the type
\beq
\delta^4 (x)\, \left\langle g^2 \Tr\left(\bar\lambda^2 \bar\lambda^2 \right)\right\rangle
\label{six}
\eeq
The single-trace operator $ \Tr\left(\bar\lambda^2 \bar\lambda^2 \right)$
was discussed in detail in \cite{Cachazo}, Sect. 2.1, where it was shown that
its vacuum expectation value vanishes. (Fully antisymmetric
in the Lorentz indices part vanishes kinematically while the
symmetric part is $Q$-exact and thus its expectation
value also vanishes\,\footnote{Note that  $ \Tr\left(\bar\lambda^2 \bar\lambda^2 \right)$ involves $d^{abc}$ structure constants.
One of the consequences
is the fact that $ \Tr\left(\bar\lambda^2 \bar\lambda^2 \right)\equiv 0$
for the SU(2) gauge group because in SU(2) there are no $d$ symbols.}.) This concludes the proof of the first line in 
Eq.~(\ref{two}).

For $n>2$ we replace the right-most operator in the $n$-point function
according to Eq.~(\ref{three}) and then drag $Q$ to the left,
generating {\em en route} various contact terms $\sim  \Tr\left(\bar\lambda^2 \bar\lambda^2 \right)$ and correlators with 
$n-2$ operators $\Tr \left( G_{\alpha\beta}\, \bar\lambda^2\right)$,
for which we repeat the substitution (\ref{three}), eventually
arriving at zero in the right-hand side.

Equations~(\ref{two}) are trivial
in perturbation theory. SUSY gluodynamics is thought to be confining,
however, with color-singlet composite states in the spectra.
At the level of color-singlet composites Eq.~(\ref{two})
implies a perfect spectral degeneracy of  hybrid mesons
with  quantum numbers
$J^{PC}=1^{+-}$ and $J^{PC}=1^{--}$, respectively,
produced by the operator $\Tr \left( G_{\alpha\beta}\, \bar\lambda^2\right)$,
much in the same way as
the vanishing of the correlators
\beq
\left\langle \Tr \left(  \bar\lambda^2 (x_n)\right) ,\, \Tr \left(  \bar\lambda^2 (x_{n-1})\right) ,\, ...\,,
 \Tr  \left(  \bar\lambda^2 (0)\right) \right\rangle_{\rm connected}
\label{seven}
\eeq
in the chiral sector \cite{NSVZ} implies a complete spectral degeneracy
in the $0^{\pm +}$ channels. The easiest way to establish quantum numbers of the operator  $\Tr \left( G_{\alpha\beta}\, \bar\lambda^2\right)$
is to rewrite it in the Majorana notation for gluino,
\beq
\Tr \left( G_{\alpha\beta}\, \bar\lambda^2\right)
\to \left( \bar \lambda \vec E\lambda + i\bar \lambda \vec B \gamma^5\lambda
\right) +i
 \left( \bar \lambda 
\vec B\lambda
-i
  \bar \lambda \vec E  \gamma^5\lambda \right)\,,
\label{deone}
\eeq
and there is no interference between the two terms. Here 
$\vec E$ and $\vec B$ stand for chromoelectric and chromomagnetic fields,
respectively. The spectral degeneracy in the
$J^{PC}=1^{\pm-}$ channels induced by
two terms in Eq. (\ref{deone})
can be viewed as a manifestation of chromoelectric/magnetic duality
appropriate to \none.

The $C$-deformed \noneh SYM theory has the Lagrangian \cite{Sei}
\beq
{\cal L}_C = {\cal L}_{C=0} +\left(\frac{1}{4g^2}\right)
\left(-i C^{\alpha\beta}\Tr \left( G_{\alpha\beta}\, \bar\lambda^2\right)
+({\rm det} C) \Tr\left(\bar\lambda^2 \bar\lambda^2 \right)
\right).
\label{eight}
\eeq
Since $Q$ is conserved for any $C$, the expansion of the vacuum energy 
${\cal E}(C)$ in
the powers of $C$ must generate zeros order by order.
This  leads to a set of the low-energy theorems
the first of which has the form
\beq
i \int d^4 x \left\langle \Tr \left( G_{\alpha\beta} \bar\lambda^2 (x)\right) ,\,
 \Tr  \left( G_{\gamma\rho} \bar\lambda^2 (0)\right) \right\rangle_{\rm connected}=0\,.
\label{nine}
\eeq
Higher order terms in $C$ emerge with higher $n$-point functions.
These low-energy theorems present a weak (integrated) form
of (\ref{two}). The validity of  (\ref{two}) locally, point by point,
means that even if one considers $C$ to be a coordinate-dependent function,
the vacuum energy density would still vanish, which, in turn,
suggests that  \noneh theories
with coordinate-dependent $C^{\alpha\beta}$ are $Q$-invariant too. 
Examination of the corresponding supertransformations\,\footnote{See Eq.~(4.16) in \cite{Sei}. The only $C$-containing term is in the
supertransformation of $\lambda$, namely, 
$(\delta\lambda_\alpha) _C = \epsilon^\beta C_{\alpha\beta}\bar\lambda^2$.
}  confirms this statement. In other words one can say
that two supercharges $Q_\alpha$ remain conserved 
in arbitrary self-dual graviphoton background, not necessarily
in coordinate-independent background. This is akin to the statement
that two out of four supercharges are conserved
in the arbitrary self-dual gauge-field background (e.g., instanton),
which leads to the Bose-Fermi spectral degeneracy in this background,
resulting in multiple exact results of the type of the NSVZ $\beta$ function
\cite{NSVZbeta}.

Now, 
let us turn to implications of the above-established spectral degeneracy
in the non-supersymmetric orientifold theory.
Using the planar equivalence \cite{ASV} we can 
pass to   orientifold theory  (at large $N$) by mere rewriting (\ref{two})
in terms of fields relevant to this theory. 
In the fermion sector we replace the Weyl adjoint gluino by a
Dirac field $\Psi$ in the two-index antisymmetric representation of
SU($N$). The gluon sector remains intact. The operator
$\bar\lambda^2$ is then replaced by $\left(
\bar\Psi\, (1+\gamma^5)\Psi\right)^i_j$ where $i,j$ are color indices,
and the trace $\Tr \left( G_{\alpha\beta} \bar\lambda^2 (x)\right) $
must be replaced by
\beq
\left( G_{\mu\nu}-\tilde{G}_{\mu\nu} \right)^j_i\,
\left(
\bar\Psi\, (1+\gamma^5)\Psi\right)^i_j \equiv {\cal G}_{\mu\nu}\,.
\label{ten}
\eeq
In the orientifold theory at $N\to \infty$ Eq.~(\ref{two}) is then replaced by
\beq
 \left\langle {\cal G}_{\mu\nu} (x_n) ,\, {\cal G}_{\rho\sigma} (x_{n-1})\,, 
 ...\,,
{\cal G}_{\phi\chi } (0) \right\rangle = 0\,.
\label{eleven}
\eeq
The consequence is the same as in \none SUSY gluodynamics:
the complete degeneracy of the meson parameters in the
$1^{\pm -}$ channels. In particular,
the spectral functions corresponding to the two-point functions
induced by the currents
$$
\left( \bar \Psi \vec E\Psi + i\bar \Psi \vec B \gamma^5\Psi
\right) 
$$
and
$$
 \left( \bar \Psi 
\vec B\Psi
-i
  \bar \Psi \vec E  \gamma^5\Psi\right)
$$
are identical.

Now, if we descend down to $N=3$, the orientifold
theory becomes one-flavor QCD \cite{ASV}. In this case we predict the spectral degeneracy in the
$1^{\pm -}$ channels up to corrections of the order of $O(1/N)$.

\vspace{3mm}

A remark is in order here regarding a straightforward generalization
of   the low-energy theorems analogous to (\ref{nine})
to SYM theories with matter. In
the  \noneh SYM  theory with  one additional flavor
(either fundamental or adjoint) one gets
an additional $C$-dependent contribution to the Lagrangian \cite{araki} of 
the form 
\beqn
\delta_C {\cal L}_{\rm matter}= \left\{
\begin{array}{ll}
-\frac{1}{\sqrt 2}C^{\alpha\beta} 
\left( D_{\alpha\dot\alpha}{\bar \phi}\right)
{\bar\lambda}^{\dot \alpha}\psi_{\beta},\,\,\, {\rm fundamental},
\\[3mm]
-\Tr\frac{1}{\sqrt 2}C^{\alpha\beta}\{D_{\alpha\dot\alpha}\bar\phi,
\,  \bar \lambda^{\dot \alpha}\}
\psi_{\beta},\,\,\, {\rm adjoint},
\end{array}
\right.
\eeqn
where $\phi,\psi$ are the scalar and spinor components of the chiral 
matter superfield  either 
in the fundamental or the  adjoint representation. Note that the deformation does not
touch the   structure of the flat directions in these theories. 

In both theories
we have two operators  coupled to the term linear in $C$ and 
a single $Q$-exact operator
coupled to $C^2$. The corresponding low-energy theorems read  
\beq
\int d^4x \langle O (x),\, O (0)\rangle_{\rm connected}= 0\,,
\eeq
where
\beqn
O= 
\left\{
\begin{array}{ll}
\frac{i}{2g^2}\Tr  G_{\alpha\beta} \bar\lambda^2 (x)
+\frac{1}{\sqrt 2}
\left( D_{\{\alpha\dot\alpha}{\bar \phi}\right)
{\bar\lambda}^{\dot \alpha}\psi_{\beta\}},\,\,\, {\rm fundamental},
\\[3mm]
\frac{i}{2g^2}\Tr  G_{\alpha\beta} \bar\lambda^2 (x)+\Tr\frac{1}{\sqrt 2}\{D_{\{\alpha\dot\alpha}\bar\phi,
\,  \bar \lambda^{\dot \alpha}\}
\psi_{\beta\}},\,\,\, {\rm adjoint},
\end{array}
\right.
\eeqn
plus similar multipoint correlators at zero momentum.
If both the adjoint and fundamental flavors are added then the 
$ C$-deformed \ntwo SQCD
emerges, where the corresponding low-energy theorems can be derived in a similar manner.

A question arises as to the possibility of
extending  the  Veneziano-Yankielowicz effective
Lagrangian \cite{VYank} to include the low-energy theorems presented above in Eq.~(\ref{nine}). This will certainly require an expansion
of the field content of the  Veneziano-Yankielowicz 
Lagrangian.  As we see,
there are nontrivial relations in the mixed holomorphic-antiholomorphic sector,
implying that additional composite fields of mixed chirality 
will have to be added.

Our last remark is of the literature character.
Low-energy theorems for the quark-quark-gluon operators of a different Lorentz structure ($J^{PC}=1^{-+}$ hybrids), were obtained in QCD {\em per SE},
with no reference to the large-$N$ limit and supersymmetry, long ago,
see Ref.~\cite{mitya}.
The current 
$J_\mu = \bar\Psi G_{\mu\nu} \gamma_\nu \Psi$ considered in these works
couples dotted and undated fermion fields, rather than the chiral fermion fields in our expressions.

The work of A.G. was supported in part by  RF BR grant RF BR-040100646. A.G. thanks FTPI of the
University of Minnesota for hospitality and support.
M.S. thanks N. Seiberg for useful comments.
The work  of M.S. was
supported in part by DOE grant DE-FG02-94ER408.

\end{document}